\documentclass[twocolumn,showkeys,preprintnumbers,amsmath,amssymb]{revtex4}

\usepackage{graphicx}
\usepackage{dcolumn}
\usepackage{bm}

\newcommand{\myemail}{john@physics.uwa.edu.au}

\begin{document}

\title{Is the Universe really expanding?}

\author{J.G. Hartnett}
\affiliation{School of Physics, the University of Western Australia\\35 Stirling Hwy, Crawley 6009 WA Australia; \myemail }

\date{\today} 

\begin{abstract}
The Hubble law, determined from the distance modulii and redshifts of galaxies, for the past 80 years, has been used as strong evidence for an expanding universe. This claim is reviewed in light of the claimed lack of necessary evidence for time dilation in quasar and gamma-ray burst luminosity variations and other lines of evidence. It is concluded that the observations could be used to describe either a static universe (where the Hubble law results from some as-yet-unknown mechanism) or an expanding universe described by the standard $\Lambda$ cold dark matter model. In the latter case, size evolution of galaxies is necessary for agreement with  observations. Yet the simple non-expanding Euclidean universe fits most data with the least number of assumptions. From this review it is apparent that there are still many unanswered questions in cosmology and the title question of this paper is still far from being answered.
\end{abstract}

\keywords{expanding universe, static universe, time dilation} 

\maketitle

\section{Introduction}
Ever since the late 1920's, when Edwin Hubble discovered a simple proportionality  \citep{Hubble1929}  between the redshifts in the light coming from nearby galaxies and their distances, we have been told that the Universe is expanding. Hubble found the recession speeds of nearby galaxies were linearly related to there radial distance with a constant of proportionality, $H_0$.  This relationship--dubbed the Hubble law--has since been strengthened and extended to very great distances in the cosmos. Nowadays it is considered to be well established in the expanding big bang universe, where the Robertson-Walker metric is used to derive a Hubble law valid for all distances. This means that the space that contains the galaxies is expanding and that the galaxies are essentially stationary in that space, but dragged apart as the universe expands--now called cosmological expansion \citep{Davis2003}. The galaxies themselves, including our solar system, do not expand as space expands in the Hubble flow \citep{Cooperstock}.

Hubble initially interpreted his redshifts as a Doppler effect, due to the motion of the galaxies as they receded for our location in the Universe. He called it a `Doppler effect' as though the galaxies were moving `through space'; that is how some  astronomers initially perceived it. This is different to what has now become accepted but observations alone could not distinguish between the two concepts. 

\citet{Zwicky} suggested a non-Doppler interpretation for the observed redshifts, an energy depletion process that has been since labeled ``tired light''. \citet{Hubble1935} considered this idea. \citet{Hubble1936} claimed his galaxy number counts supported this type of linear energy depletion but he could offer no plausible mechanism. Later in his life \citet{Hubble1947}  said that the Hubble law was due to a hitherto undiscovered mechanism, but not due to expansion of space. See \citep{LaViolette1986} for a discussion on ``tired light''.

The fact that the Hubble law can be derived from general relativity, which has been successfully empirically tested in the solar system by numerous tests \citep{Reynaud2008}, and with pulsar/neutron star and pulsar/pulsar binary pairs \citep{Hulse1975, Burgay2003, Lyne2004} in the Galaxy, is a very strong point in its favor, and strong evidence for an expanding universe \citep{Peacock1999}. But it does not prove it, and, unless a physical mechanism can be established that produces a Hubble law in a static universe then this fact favors the expanding universe. 

However, to date, there is no experimental local laboratory evidence that establishes cosmological expansion as a real phenomenon of nature \citep{Carrera2006}.  Though it can be derived as a consequence of Einstein's general relativity theory,  it has been claimed by some as a fudge factor  \citep{Lieu2007}  to support what has now become the standard model--the  $\Lambda$ cold dark matter ($\Lambda$CDM) model, also called the concordance model. As observations were made at larger and larger scales, the standard model has required the introduction of exotic non-baryonic ``dark'' matter and ``dark'' energy ($\Lambda \neq 0$) providing a sort of anti-gravity. 

Therefore, it seems opportune at this moment in time to review the evidence both for and against the concept of cosmological expansion. As an alternative, it is necessary to compare the expanding universe to a static universe. Though no inference should be drawn on the author's personal view here.

\section{The physical evidence}

All evidence for cosmological expansion comes from the cosmos itself. ``\textit{The cosmological redshift is important because it is the most readily observable evidence of the expansion of the
universe.}'' \citep{Davis2003}. 

Supernovae are among the brightest light sources in the sky. Astrophysicists believe that they have successfully understood the origin of a certain class of these explosions using general relativity, where a white dwarf star, after accumulating sufficient mass from a companion star to reach the critical Chandrasekhar mass limit, catastrophically collapses in on itself under its own gravity and explodes in a blinding flash of light. The luminosity of the explosion rapidly increases, peaks, and then slowly decreases over days and months. By modeling this it is believed that one can understand what the intrinsic brightness at the peak of the explosion was and hence one can establish, for a certain class of these supernovae, a `standard candle'. The theory says that the intrinsic brightness at the peak of the explosion is the same for all supernova in this class--the type Ia, which are identified from the metal content in their spectra. Hence if you know their intrinsic brightness you can determine their distance in the cosmos. Then using the redshifts of their host galaxies, the distance modulus, derived from the standard cosmology,  can be tested with the matter density ($\Omega_m$),  the dark energy density ($\Omega_{\Lambda}$)  and the Hubble constant ($H_0$) as the only free parameters \citep{Perlmutter}. 

From this method it has been determined not only that the Universe is expanding but also that the expansion is accelerating \citep{Riess1998}.  In order for  the observations to fit the standard cosmology it has been necessary to add dark energy with a non-zero value for the cosmological constant ($\Lambda$) and also a significant amount of dark matter. And it follows that together these comprise about 96\% of the mass-energy content of the Universe. Without them the  $\Lambda$CDM big bang (BB) model seriously fails to describe the observed luminosities.  Besides dark energy and dark matter are totally unknown entities in the laboratory. Though enormous effort has been made to detect putative dark matter particles from the Galactic halo  all efforts have so far failed \citep{Aprile2010, Aprile2011}.

One of the consequences of cosmological expansion is \textit{time dilation}. When the light curves, which show the rise and fall in luminosity of the supernova explosion, are compared at increasing redshifts their time axes should be stretched due to time dilation with respect to the observer at the Earth. In other words, processes that follow a flow of time in the distant cosmos are slowed relative to Earth time. Such a ``time dilation'' effect has been clearly observed in the light curves of the type Ia supernovae and is claimed as definitive evidence for expansion  \citep{Riess1997, Goldhaber}. Yet, no time dilation has been observed in the luminosity variations of quasars  \citep{Hawkins2001, Hawkins2010},  which are meant to be at very great distances based on their redshifts and the Hubble law. How can these contradictory claims be reconciled?

Add to this  evidence suggesting that some quasars are apparently associated with relatively low redshift galaxies  \citep{ArpQRC, ArpSR, Bell2002, Galianni, ArpFulton},  which can only be reconciled if those quasars are not at their redshift distances but are located nearby.  And the fact that \textit{proper motion}  is observed in quasars  \citep{Varshni, Talbot, MacMillan}  really brings into doubt that at least some of them must not be at the cosmological distances derived from their redshifts. That means that a large part of a quasar's redshift must be due to some as-yet-unknown non-cosmological cause, i.e. not due to expansion of space. If verified this is very damaging to the standard model. And considering that quasars in the standard model are at cosmological distances, they should be young objects. Their larger redshifts imply younger quasars. Therefore, quasars should be deficient in metals at higher redshifts, which  should be observed in their metal abundances as a function of epoch. But observations show no  metal deficiency as a function of redshift \citep{Fan, Simon}. Quasar environments, based on their emission lines, are generally metal rich with metallicities near or above the solar value even to the highest redshifts.

Considering the history of the expanding universe hypothesis, the burden of proof should really rest with those that make the claim. Hubble first thought that the redshifts of the galaxies was due to a Doppler effect (motion of the galaxies through space) but as cosmology developed it was shown theoretically that the effect could be understood as resulting from the expansion of space over the period of flight of the photons from emitter to receiver. And the reality is it is claimed to be independent of the emitter source. If independent then that means the origin of the redshifts comes from a process during the flight of the photon from source to receiver. The expansion of space itself is the best argument currently for this.

The question must be asked, what physical evidence do we have that the universe is expanding?   \citet{Lopez-Corredoira2003} reviewed the evidence for this and other questions for cosmology today. This paper reiterates and updates the review of some of those same lines of  evidence. So besides the redshifts themselves what evidence exists.

\section{Evidence for time dilation}
\subsection{Type Ia supernovae}
The type Ia supernova (SN) measurements are the very best evidence for an expanding universe. In 1998 two independent projects (the Supernova Cosmology Project  and the High-z Supernova Search) confirmed that the Universe was expanding but also announced that it was accelerating  \citep{Riess1998}.  The supernova  light-curve peak luminosity ($L$) was correlated to an absolute magnitude ($M_B \propto -2.5 log(L)$), which is assumed to be  intrinsic  to that class of supernovae. 

The light curves were adjusted for a stretch factor $w = s(1+z)$ which is claimed to be due to time dilation as a function of epoch ($z$), the redshift of the source. This is absolutely required in an expanding universe. In fact, it is the only redshift mechanism on offer that requires it. To my knowledge this time dilation factor is the only evidence for an expanding universe that sets it apart from a static universe. The Hubble law, or the relationship between the apparent magnitudes and redshifts of galaxies, is not sufficient grounds to establish an expansion.  Since \citet{Zwicky} proposed his \textit{tired light} idea many other possible redshift mechanisms  have been theoretically suggested, though none have gained any sort of general acceptance like cosmological expansion has. To date one author has compiled 31 mechanisms giving a quantitative description of how large redshifts may possibly be related to distance \citep{Marmet}.

With the analysis of the supernova light-curves the stretch factor ($w$) correction is determined by hand, an empirical fit to the best selected data.  The study that showed the most constrained results found a sample of light curves proportional to $(1+z)^b$ where $b = 1.07 \pm 0.06$  \citep{Goldhaber}.  This seems to be the most definitive measurement of time dilation where $b$ should be identical with unity. However, a possible criticism is that the time under the light-curve could depend on the intrinsic brightness of the supernovae (i.e. the correction factor $s$), which might vary considerably with the redshift.   \citet{Lopez-Corredoira2003} provides a very good review of this. 

A similar point is made by  \citet{Crawford2011}: \textit{``Since current investigators assume that the type Ia supernovae have essentially a fixed absolute BB magnitude (with possible corrections for the stretch factor), one of the criteria they used is to reject any candidate whose predicted absolute peak magnitude is outside a rather narrow range. The essential point is that the absolute magnitudes are calculated using BB and hence the selection of candidates is dependent on the BB luminosity-distance modulus.''}

Basically he is claiming \textit{selection bias.} Is not this circular reasoning? If you select only the candidates that fit the desired luminosity-distance criteria and use them to determine the luminosity distance.  Since one cannot determine the absolute magnitudes of the sources without assuming a cosmology, the standard concordance criteria ($\Omega_m \approx 0.3$,  $\Omega_\Lambda \approx 0.7$, and $H_0 \approx 70$ km/s/Mpc) are used to calculate the absolute magnitudes for the candidates, which must be in a narrow range near $M_B \approx -19$, and the acceptable ones are used to test the same model, and therefore determine values for $\Omega_m$ and $\Omega_\Lambda$. This is confirmed by \citep{Foley} who state \textit{``...for any individual SN Ia, the intrinsic width is unknown, so without assuming a $(1+z)$ dilation, the intrinsic width and dilation cannot be separated.''}

Nevertheless for the selected supernovae \citep{Goldhaber} the regression fit to the derived absolute magnitudes ($M_B$) of the sources on the expected $2.5 log(1 + z)$ redshift dependence shows that the luminosity is proportional to $(1 + z)^a$ where $a = 0.23 \pm 0.07$. This means that their intrinsic luminosity must have slowly decreased as the universe evolved.  There is no reason why the mass of the white dwarf progenitor stars for these supernovae should increase as the Universe ages, hence resulting in brighter explosions. One of the assumptions of the Cosmological Principle is that the physics of the Universe is the same at all epochs.    Note Fig. 13 (page 1036) of \citep{Riess1998}  where various SN Ia light curves are shown with different absolute magnitudes $M_B$. The brighter sources decline slower than the dimmer sources. The standard explanation for this change is the \textit{ad hoc introduction of dark energy}  \citep{Turner}  or \textit{quintessence}  \citep{Steinhardt}. Hence evolution in the size and mass of the galaxies over cosmic time has been assumed as the reason. The question then remains what level of circular reasoning has been used for selection of the candidate type Ia supernovae  because they do not (as initially assumed for a `standard candle') have the same intrinsic luminosities?

\citet{Crawford2011} models the luminosities of type Ia supernova in a static universe and finds that the total energy of the explosion (area under the light curve) is a far better `standard candle'.  Therefore assuming that all these type of supernova have the essentially the same energy, based on the modeling of the critical Chandrasekhar mass limit of the progenitor white dwarf, the product of the peak luminosity and the width of light curve will be a constant.  Since the prime characteristic used for selecting these supernovae is the peak absolute magnitude, which is computed using the standard concordance model,  there is a strong bias that results in intrinsically weaker supernovae being selected at higher redshifts. The absolute magnitudes  cannot be determined without assuming a cosmological model first.  And for constant energy the weaker supernovae must have wider light curves. This is a selection effect that has width of the light curve increasing with redshift and hence can mimic time dilation in the resulting selected candidates.

When \citet{Crawford2011} applies his model of absolute energy (absolute magnitude in his static model plus correction for width) for each supernova in the same SN Ia data sets  \citep{Uniondata}  used to test the standard model he finds the energy of the explosion to be invariant over all redshifts with a curve-fit slope of $0.047 \pm 0.089$, which is consistent with zero. This means no change over all redshifts. Using a simple selection model for SN Ia data he shows their width dependence on redshift, and considering the biased nature of the data, is a very reasonable fit. Hence no time dilation and no cosmological expansion. Because no additional energy is needed for the fit, no dark energy or quintessence is needed either. 

In an effort to resolve this \textit{time dilation} question in supernova light-curves a single supernova (1997ex) was studied  \citep{Foley}  at different epochs separated by months and found that the spectral evolution of the source is inconsistent with no time dilation at a 96.4\% confidence level. The claim lies in the spectral-feature age that is used to independently determine the aging of the source at approximately monthly intervals. The derived age measure is then compared to the expected $(1+z)$ aging. Hence the amount of aging in the supernova rest frame should be a factor of $(1+z)^{-1}$ smaller than that in the observer frame. The results were found to be consistent with time dilation. 

It should also be mentioned that this latter paper discusses the consistency of time dilation seen both in the SN light-curve, over monthly timescales, and in the wavelengths of the light seen in the observer frame, i.e. in the redshifting of the light from the source. This is the important distinction for this review. Are longer timescale time measures consistent with the ``\textit{femtosecond time dilation}'' in the observed redshift of the light from the sources? 

The concept of the accelerating universe has come from the very highest redshift type Ia supernova observations, and hence the idea of dark energy (or a cosmological constant) driving the Universe apart. This has resulted from a deficit of the expected luminosity determined from the standard model with $\Lambda = 0$ and that observed in these distant sources. However it has also been criticized on the basis of intergalactic dust  \citep{Aguirre, Goobar}, causing the added deficit and that the presence of grey dust is not inconsistent with the measure on the most distance supernova at redshift $z = 1.7$ (SN 1997ff)  \citep{Goobar}. 

Type Ia supernovae may also have a metallicity  dependence on redshift which may mean that the resulting non-zero value of the cosmological constant may require corrections for metallicity by factors as large as the effects of the assumed cosmology itself  \citep{Rowan-Robinson}.  This causes an underestimate of the effects of host galaxy extinction; a factor which contributes to the apparent faintness of the high redshift supernovae is evolution of the host galaxy extinction as a function of redshift, caused by the presence of molecular clouds and dust. Therefore with a proper treatment of the latter, and if one eliminates those SN Ia sources not observed before peak brightness is reached, the evidence for a cosmological constant (and dark energy) is quite weak.

The use of standardized SN Ia light curves involves the stretch parameter ($s$) \citep{Perlmutter} related to both the width of the light curve and the magnitude at maximum brightness. This determined empirically from observational data and based on events at low redshift, and only assumed to be valid at high redshift. \textit{``...if a systematic different relation holds for high-z events (either in the average value, or in the dispersion, or both) the cosmological application of SNe Ia as distance indicators would be called into question.''} \citep{Greggio2010} 

As a result a concerted effort is being made to understand what the SN Ia progenitor stars are. This is still being debated  \citep{Greggio2010, Wang2010, Howell2009}. How does metallicity affect the mass of the progenitors and hence the SN Ia luminosities? There is no clear resolution as yet. \citet{Howell2009} state that \textit{``Age may have a greater effect than metallicity--we find that the luminosity-weighted age of the host galaxy is correlated with $^{56}$Ni yield, and thus more massive progenitors give rise to more luminous explosions. This is hard to understand if most SNe Ia explode when the primaries reach the Chandrasekhar mass. Finally, we test the findings of \citep{Gallagher2008} that the residuals of SNe Ia from the Hubble diagram are correlated with host galaxy metallicity, and we find no such correlation.''} That is, the dispersion in the distance modulii  from those expected is due to difference in the metal content of the SN Ia environments.

It is worth noting that there have been various attempts to construct alternate models that fit the SN Ia data without time dilation.  \citet{Ivanov2001} has developed a quantum gravity static universe model   that has a Hubble law resulting from quantum interactions. There is no time dilation in his model. The author compares the predictions of his model with both SNe Ia and GRBs without time dilation  \citep{Ivanov2010}.  He corrects the published SN Ia distance modulii for the time dilation stretch factor and compares with his model. The fits are extremely good yet no dark energy term is needed. \citet{Ivanov2010} concludes his paper with the telling remark, \textit{``...the discovery of dark energy in a frame of the standard cosmological model is only an artefact of the conjecture about an existence of time dilation.''}

One can say then that if there exists at least one static model where if one corrects the SN Ia data for no time dilation and it fits that model then that creates significant doubt about the need for dark energy and dark matter in the first instance.

\subsection{Quasar luminosity variations}
 Quasars show variations in their luminosities over timescales of weeks to years. This means that quasars generate and emit their energy from a very small region, since each part of the quasar would have to be in causal contact with other parts on such time scales to coordinate the luminosity variations. As such, a quasar varying on the time scale of a few weeks cannot be larger than a few light-weeks across. 
 
 And to date from extensive observations of quasars no time dilation has been found in their luminosity variations \citep{Hawkins2001, Hawkins2010}. \citet{Hawkins2010} used the light curves of over 800 quasars monitored on time scales from 50 days to 28 years. He divided his data into two groups, for quasars at low ($z < 1$) and high redshift ($z > 1$) and used Fourier power spectral analysis methods. He compared their spectral energy distributions (SEDs), at high and low redshifts, to look for changes expected from time dilation. The research has found that the SEDs at high and low redshift are identical.
 
 The research also confirmed   an anti-correlation  observed between the luminosity and the amplitude of the light curves of the quasars. For a sample of quasars, the more luminous are seen to vary over a smaller range of brightness than the less luminous ones. It would seem then that this fact would make it difficult to resolve a time dilation effect. But Fourier analysis provides a way of giving a measure of the variability on different time scales and separate it from magnitude effects. With a sufficient time span of the data the degeneracy between time-scale and amplitude (magnitude) can be resolved.  The results of this research is powerful evidence against any time dilation effects in the Universe as a function of epoch. 
 
 If the quasar do not show time dilation when considered from the observer's frame of reference, how can these measurements be reconciled with the SN Ia measurements? \cite{Hawkins2010} discusses possible explanations to compensate for the lack of time dilation  and involves the possibility that time dilation effects are exactly offset by an increase in the timescale of variations associated with black hole growth (that is thought to power the quasar), or that the variations that are observed are caused by microlensing \citep{Hawkins2007}, hence not intrinsic to the quasar. The latter means the variations do not originate in the quasars themselves but along the line of sight at lower redshifts. In such a case time dilation would not be expected. But these would have to occur in the same manner over all timescales. Such explanations are not very satisfactory.
 
 One possible resolution  is that the quasars are not at the cosmological distances indicated by their redshifts but are in fact much closer \citep{ArpSR, Galianni, ArpFulton}. Of course, the implication is that some quasars are in some way different from galaxies (at least the mechanism generating their redshifts is) but are associated with low redshift galaxies \citep{Lopez-Corredoira2007, Arp2001}.

\subsection{GRB luminosity variations}
Gamma-ray bursts (GRBs) are flashes of gamma rays associated with extremely energetic explosions observed in the distant cosmos \citep{Piran2004}. Assuming their host galaxy redshift are a good measure of distance, they are the most luminous sources in the universe. Bursts can last from ten milliseconds to several minutes. The initial burst is usually followed by a longer-lived ``afterglow'' emitted at longer wavelengths, covering all parts of the electromagnetic spectrum (X-ray, ultraviolet, optical, infrared, and  from microwaves  to radiowaves). Their peak energies are in the gamma ray and X-ray parts of the spectrum. Most observed GRBs are believed to consist of a narrow beam of intense radiation emitted from a supernova \citep{Bloom2003}. 

The claim has been made that GRBs show time dilation in time measures of the gamma ray bursts \citep{Chang2001, Chang2002, Borgonovo2004}. \citet{Chang2001} attributed the anti-correlation of one time measure with a brightness measure indirectly as evidence of time dilation itself. \citet{Norris2002} and \citet{Bloom2003} claim that this is the reason why the time dilation cannot be observed in the raw data. Because a strong luminosity dependent selection produces an average luminosity that increases with redshift there is a simultaneous selection of time measures that decrease with redshift and this cancel the observable effects of time dilation. 

\citet{Shen2003} found a bimodal distribution of GRBs where the long GRBs are composed of two sub-classes with different time variability in a time measure, the power density. Their claim is that the averaged variability time scale decreases with the peak flux and is consistent with the expected time dilation. 

But Hawkins (2010) states that the evidence for time dilation from gamma ray bursts is inconclusive. Initially, that was because of the uncertainty in the intrinsic timescales of the bursts, but later, once the redshifts of bursts were found, the problem of correcting the raw data for selection effects involving an inverse correlation between luminosity and time measures made it difficult to use GRBs to detect time dilation.  

Four time measures, determined from the original gamma-ray observations, are independent of any model for the burst mechanism \citep{Schaefer2007}.  The relevant measures include the lag time between a band of high energy gamma-rays and a band of lower energy gamma-rays ($\tau_{lag}$), the shortest time over which the GRB light curve rises by half the peak flux of the pulse ($\tau_{RT}$), and the number of spikes or variations per second in the light curve ($V$), which is estimated with respect to a smoothed version of the light curve. The fourth is  the  estimated time span that contains 90\% of the counts ($T_{90} $). 

\citet{Crawford2009} makes a careful analysis of the traditional explanation that an inverse correlation between luminosity and these time measures together with strong luminosity selection as a function of redshift cancels any observed time dilation. He confirms that there is an inverse correlation between luminosity and some time measures. Of the 4 listed above it is strongly seen in 2 of them. But using the concordance cosmology strong luminosity selection cannot be achieved. He  finds that GRBs out to $z = 6.6$ show no evidence of time dilation in the raw data and rejects the hypothesis with a probability of $4.4 \times 10^{-6}$. 

It may be possible to explain the apparent lack of time dilation with a combination of gamma-ray burst selection, some luminosity evolution and some time measure evolution. But \textit{this requires a remarkable coincidence, where opposite effects exactly cancel, in order to produce the apparent lack of time dilation}. However the data are consistent with a static cosmology in a non-expanding universe. \citet{Crawford2009} finds that, assuming a static universe, the total energy of the GRBs is found to be invariant with redshift. This is a similar result that can be shown in the type Ia supernova data also.

\section{Evidence against expansion}
\subsection{Angular size test}
The test of the dependence of the angular size of some sources with redshift was first conceived by  \citet{Hoyle}.  In principle, it is simple, but in application not so simple, because of the difficulty in finding a `standard rod', a type of object that undergoes no evolution in linear size over time spans of order of the age of the Universe. The angular sizes of quasars (or quasi-stellar objects (QSOs)) and radio galaxies at radio wavelengths, for first ranked cluster galaxies in the optical, and for the separation of brightest galaxies in clusters or in QSO-galaxy pairs of the same redshift have all been measured.   \citet{Lopez-Corredoira2010}  provides an excellent analysis of this and the Tolman surface brightness test. See also the references contained therein.

This type of test is related to the Tolman surface brightness test but tests for the angular size ($\theta$) of an object as a function of epoch ($z$).  These will vary quite differently depending on the cosmology assumed. The angular sizes of radio galaxies over a range up to $z = 2$ show a dependence $\theta \propto  z^{-1}$  \citep{Andrews,Kapahi}, which is a static Euclidean effect over all scales.  Size evolution as a function of redshift is needed for this to fit the standard model.

In the standard model evolution in object size is assumed ad hoc and generally is used to make up for any deficiency been the modeled and observed size as a function of redshift. Any discovered $\theta \propto  z^{-1}$ dependence, as predicted by a static Euclidean universe, would be just a fortuitous coincidence of the superposition of the angular size $\theta(z)$ dependence in the expanding universe with evolutionary and/or selection effects. However, the fit of radio source counts was found to be best when no evolution was assumed  \citep{DasGupta}.    \citet{Lopez-Corredoira2010} found that, when assuming the standard cosmological model as correct, the average linear size of galaxies, with the same luminosity, is six times smaller at $z = 3.2$ than at $z = 0$, and their average angular size for a given luminosity is approximately proportional to $z^{-1}$. 

Neither the hypothesis that galaxies which formed earlier have much higher densities nor their luminosity evolution, nor their merger ratio, nor massive outflows due to a quasar feedback mechanism are enough to justify such a strong size evolution. Without a very strong size evolution the standard model is unable to fit the angular size vs. redshift dependence. This requires between 2 and 4 major mergers per galaxy during its lifetime, which is observationally unjustifiable. Also it is not known how local massive elliptical galaxies have grown as similar sized galaxies are known at high redshifts. Therefore it follows that the nearby ones must have been much smaller at high redshift assuming size evolution to be true. And no method is known how spiral galaxies grow through mergers and preserve their spiral disk nature. 

Some disk galaxies have been found that have no nuclear bulge; they are considered to be almost too good to be true \citep{Kormendy}. \citet{Kormendy} ask the question:\textit{``How can hierarchical clustering make so many giant, pure-disk galaxies with no evidence for merger-built bulges?''} Simulations show as spirals merge their spiral disk structure is lost. Seems like no mergers have occurred with these galaxies over their lifetimes. And observations of five brightest cluster galaxies (BCGs) at redshifts $0.8 < z < 1.3$ were compared to a group of BCGs at $z = 0.2$ and they were found to be no more than 30\% smaller indicating little or no evolution has occurred, contrary to the standard model  \citep{Stott}. 

However, in a study \citep{Shim} of 74 galaxies with spectroscopic redshifts in the range of $3.8 < z < 5.0$ over the Great Observatories Origins Deep Survey (GOODS) fields evidence is presented for strong H$\alpha$ emission. This is inferred from an excess of 3.6$\mu m$ radiation. The strong H$\alpha$ emission then implies a strong sustained star formation phase over the life of the galaxies where at least 50\% of the stellar mass is accumulated at a constant rate assuming the gas supply is sustained. This is suggested to be the case in 60\% of the H$\alpha$ emitters. So early galaxy formation was not dominated by mergers but by smaller galaxies supersizing themselves by gobbling up surrounding fuel, creating an unusual amount of plump stars, up to 100 times the mass of our sun. If this model proves to be true it could get around the merger problem.

As mentioned, the main difficulty with this type of measure is establishing the standard size of the objects being observed. However, the cosmological model that uses a very simple phenomenological extrapolation of the linear Hubble law in a Euclidean static universe fits the angular size vs. redshift dependence quite well, which is approximately proportional to $z^{-1}$. There are no free parameters derived ad hoc, although the error bars allow a slight size/luminosity evolution. The type Ia supernovae Hubble diagram can also be explained in terms of this static model with no ad hoc fitted parameter, i.e. no dark matter nor dark energy. 

\subsection{Tolman surface brightness}
\citet{Hubble1935}  proposed the so-called Tolman test based on the measure of the brightness of galaxies as a function of epoch. A galaxy at redshift $z$ differs in surface brightness depending on whether there is recession or not. The units chosen for magnitude determines the redshift dependence and in bolometric units the surface brightness of identical objects in an expanding universe varies by $(1+z)^4$: one $(1+z)$ factor due to time dilation (a decrease in photons per unit time), one factor $(1+z)$ from the decrease of energy per photon and two factors from the fact that the object was closer to us by $(1+z)$ when the light was emitted. In an expanding universe regardless of the units the ratio of surface brightness in an expanding and non-expanding universe is $(1+z)^{-3}$. This is independent of wavelength.

\citet{Lerner2006}  tested the evolution of galaxy size hypothesis that is used to fit the standard model to the observed angular size of galaxies as a function of redshift. His method is based on the fact that there is a limit on the ultraviolet (UV) surface brightness of a galaxy, because when the surface density of hot bright stars and thus supernovae increases large amounts of dust are produced that absorb all the UV except that from a thin layer. Further increase in surface density of hot bright stars beyond a given point just produces more dust, and a thinner surface layer, not an increase in UV surface brightness. Based on this principle, there should be a maximum surface brightness in UV-rest wavelengths independent of redshift. \citet{Scarpa}  measured, in low redshift galaxies, a maximum FUV (155 nm at rest) emission of 18.5 $mag_{AB}/arcsec^2$ and no galaxy should be brighter per unit angular area than that. \citet{Lopez-Corredoira2010} using data from  \citep{Trujillo}  determined surface brightness values for galaxies under the assumptions of both expanding and static universes. They found that in the expanding case many galaxies would have to be brighter than the allowed limit by even up to 6 times. In the case of the static universe no galaxy would be brighter than this limit. In addition it has been reported for clusters $z> 1$ that they also are found to be ``too big, too early'' if the parameters of the standard concordance model are used \citep{Hoyle2011}.

\citet{Lerner2009}  using a large UV dataset of disk galaxies in a wide range of redshifts (from $z =0.03$ to $z=5.7$) which included 3 sets of galaxies at low redshift ($z \leq 0.1$) and 8 sets of galaxies at high redshift ($0.9 < z < 5.7$) from the Hubble telescope Ultra-Deep Field show that there is a decided preference for a fit to the angular size data with a Euclidean non-expanding (ENE) universe over that of the expanding  $\Lambda$CDM concordance model. In fact, the results are \textit{a very poor fit to the  $\Lambda$CDM model}. If the redshift range is restricted to $0.03 < z < 3.5$ then \textit{the ENE model provides a reasonably good fit}. When a very small amount of extinction is allowed for the fit is near perfect. 

\subsection{The CMB radiation}
There are two important issues here in relation to an expanding universe. 
\begin{enumerate}
	\item Can we really be sure that the \textit{cosmic microwave background radiation} (CMBR) is from a background source, that it is relic radiation from the big bang?  
	\item Does measurement of the temperature of that radiation at different epochs tell us something cosmological? 
\end{enumerate}
		
\citet{Gamow} predicted relic radiation from the big bang and thus the CMBR was a successful prediction of the standard model  but unless you could show it could not originate elsewhere it would not be proven. \citet{Lieu2006} showed that when 31 relatively nearby clusters of galaxies (where most $z < 0.2$) were studied for any decrement in temperature, a shadowing of the CMBR by the clusters, it was only detected in about one quarter of the clusters. They looked for the expected temperature decrement of the X-ray emitting intergalactic medium via the Sunyaev-Zel'dovich effect (SZE) and found sometimes even a heating effect. \citet{Bielby} extended that work in 38 clusters to show that not only was the SZE less than what was expected but that it tendered to progressively disappear for redshifts from $z=0.1$ to $z=0.3$. Their result is statistically equivalent to a null result (no shadowing) at about the 2$\sigma$  level. 

This then brings into doubt the fact that the CMBR is from the background, i.e. from the big bang and therefore whether cosmic expansion is a valid hypothesis. However to examine that more precisely one should study the temperature of this radiation at past epochs.

\citet{McKellar}  interpreted interstellar absorption lines in the blue part of the optical spectrum arising from diatomic CN molecules as being excited by background radiation with a blackbody spectrum and a required temperature of $2.3$ K. This was from sources in the Galaxy and well before ``the discovery'' of the CMBR.  

The standard cosmology predicts that the temperature of CMBR scales with redshift and that the temperature is higher than that in the solar system by the factor $(1 + z)$. Hence from the excitation of atomic transitions in absorbing clouds  at high redshifts along the line of sight to distant quasars, assuming the atoms are in equilibrium with the CMBR, this temperature can be determined. In one such case  \citep{Songaila}  a temperature of $7.4 \pm 0.8$ K at $z = 1.776$ was derived, which agrees very well with the theoretical prediction of $7.58$ K. However, another component of the same cloud, with a very similar redshift, yielded a temperature of $10.5 \pm 0.5$ K, not in such good agreement with the standard cosmology. Others also found a similar result  \citep{Ge}.  And measurements on a cloud at $z = 2.34$ resulted in temperatures between $6$ K and $14$ K  \citep{Srianand}.  This is in accord with the $9.1$ K predicted by the standard cosmology but with larger errors. 

From the analysis of the C+ fine-structure population ratio in the damped Lyman alpha (Ly$\alpha$)   absorber system towards a quasar   at $z = 3.025$ \citep{Molaro} a temperature of $14.6 \pm 0.2$ K was calculated, compared to the  theoretical prediction of $10.97$ K. The discrepancy is attributed to the existence of other mechanisms of excitation, like collisions for example. But that means that other measurements (in other papers) should also be affected by other mechanisms of excitation and that means those measurements can just give the maximum CMBR temperature, but not the minimum. \textit{Are we expected to believe that when the results agree with the theoretical predictions, no other mechanisms are involved, but when the results do not agree, they are?} 

Therefore, it seems that the increase of CMBR temperature as a function of redshift ($z$) by the factor $(1 + z)$ has not been proven. It is still an open question that \citet{Lamagna} have suggested a method to answer. 

\subsection{Absorption systems and Ly$\alpha$  lines}
When neutral hydrogen (H1) clouds are illuminated by a quasar in their background, absorption lines are seen at redshifts less (shorter wavelengths) than that of the quasar. These result from the fundamental Lyman excitation of the neutral atoms, from around 121.6 nm (for Lyman alpha, Ly$\alpha$) to 102.5 nm (for Lyman beta, Ly$\beta$). They are found in the vacuum ultraviolet part of the spectrum. The presence of a very large group of these lines, called the Ly$\alpha$ forest, representing many foreground hydrogen clouds, has been said to be a very good probe of the intergalactic medium  \citep{Rauch}.  

The Ly$\alpha$  forest seems to be very good evidence that the quasars are at their large redshift distances. It would seem to contradict the claim of Arp and others that some quasars have large intrinsic redshifts that are not due to cosmological expansion. The light from the quasar is uniformly redshifted. If this is due to some intrinsic effect it would not translate into a series of lines representing lower and lower redshift distances towards the observer from absorbing hydrogen clouds in the foreground of the quasar. The absorption lines are measured at redshifts less than that of the quasar hence the clouds should be at their cosmological redshift distances in an expanding universe.

However, all is not as it might first appear.  \citet{Prochter}  published observations that they described as `astonishing'. They found by using spectra of GRBs they were able to \textit{``... identify 14 strong MgII absorbers along 14 GRB sight lines (nearly every sight line exhibits at least one absorber)...''}. This meant that every GRB they observed showed at least one absorbing cloud/galaxy in its foreground, whereas only one quarter of quasars show the presence of absorbing clouds/galaxies. 

What is so special about GRBs that they always have an absorber in their foreground? This was discussed in a letter to the journal \textit{Science}  \citep{Schilling}  where it was mentioned that these features observed in the GRB spectra might be intrinsic to the `home galaxy' that hosts the gamma-ray burst and not to foreground galaxies. In the case of this study they used MgII lines and not H1 lines.

Lanzetta of Stony Brook University in New York is quoted by the \textit{Science} article,
\textit{``If I had to bet, I would say this is that one-in-10,000 statistical fluke that happens every now and then,'' ... ``It will probably go away when more observations become available. We'll have to wait and see.''  If the puzzle remains after 15 or 30 more GRBs are analyzed, however, then ``something very strange must be going on.'' }

By 2009, \citet{Tejos}  found that the number of absorbing systems towards GRBs was three times larger than towards quasars (from a sample of 8 GRBs studied) and no good explanation for the anomaly is forthcoming, though a few have been proposed. This then adds doubt to the proposition that the Ly$\alpha$  lines represent neutral hydrogen clouds, absorbers, in the foreground of the quasars also.

A Gunn-Peterson trough is believed to result when many Ly$\alpha$  absorption lines overlap due to many clouds of neutral hydrogen. This is theorized to have occurred towards the end of the era of reionization. The Gunn-Peterson trough is seen in the spectra of some quasars, and is strongly dependent on redshift. It is not seen in all quasar spectra. The standard model explains this where the intergalactic medium has been re-ionized--hence no absorption. The Gunn-Peterson trough is evidence for the era of the dark ages (high opacity) where there is only neutral hydrogen.

\citet{Lopez-Corredoira2003} describes some observations on this.
\textit{``A hydrogen Gunn-Peterson trough was predicted to be present at a redshift $z = 6.1$ \citep{Miralda-Escude}.    Indeed, a complete Gunn-Peterson trough at $z = 6.28$ was discovered  \citep{Becker}, which means that the Universe is approaching the reionization epoch at $z_r = 6$. However, galaxies have been observed at $z = 6.68$  \citep{Lanzetta},  or $z = 6.56$ without the opacity features  \citep{Hu}  prior to the reionization, and the epoch of reionization was moved beyond $z_r = 6.6$  \citep{Hu}}.

\textit{An inhomogeneous reionization  \citep{Becker} is a possibility to explain the apparent disagreement of the different data. Recent measures of CMBR anisotropies by the WMAP observations give a reionization epoch $z_r=20_{-9}^{+10}$ (95\% CL).  \citep{Bennett2003}  If we were going to believe that CMBR anisotropies are being correctly interpreted in terms of the standard cosmology, we would have again a new inconsistency.''}

This means that the data and the theory do not really coincide. A Gunn-Peterson trough is observed at a redshift well after the epoch $11 < z < 30$ for the era of reionization determined from CMBR observations. So is it really due the theorized effect?

For the hydrogen cloud absorption lines to show a large redshift and for the latter not to result from cosmological expansion then those lines would have to originate in the atmosphere of the quasar and be generated by the same unknown intrinsic effect as that of the quasar. As the light passes through a quasar's atmosphere the H1 atoms, as a function of distance above the quasar, would have to have different Doppler speeds inward and hence slightly less redshifted than the putative parent quasar. In other words, it has to be some mechanism connected to the quasar itself. If not, the standard model has a good argument in favor of cosmological expansion.

\citet{Ashmore2009}   reviewed and analyzed the spacing of hydrogen clouds as a function of redshift, by taking literature data on numbers of neutral hydrogen clouds measured as a function of redshift from their absorption lines with background quasars. He made the usual BB assumptions that quasars are at their redshift distances and that the Ly$\alpha$  absorption lines result from hydrogen clouds in the foreground of quasars.

From this \citet{Ashmore2009} showed that the cloud spacing is constant out to a redshift of $z \approx 0.5$ when most studies are combined and out to $z = 1.6$ from one particular survey \citep{Kirkman}. Beyond $z \approx 0.5$ generally there is a decrease in cloud spacing from other studies. With standard assumptions this would mean the Universe expanded up to $z \approx 0.5$ and then became static. If it once expanded, it describes an expanding universe that decelerated and became static. Here there is the assumption for a sort of generic static model that has no redshift dependence on line spacing but that is not necessarily the case. The specifics of a particular static cosmology may require otherwise.

Also the Doppler line broadening from the clouds indicates a near linear decrease in temperature as a function of redshift, which is the opposite of what one expects from the standard model. Above the increased redshift dependence on the temperature of the CMBR was discussed. However, if the temperature determined from the H1 line broadening is indicative of the intergalactic medium then it implies that the CMBR must be local. For a perfect black body spectrum if the CMBR arose from the earliest times it must have begun at a lower temperature than observed locally. Certainly, within the constraints of the standard cosmological model these observations are contrary to what would be expected. 

Mainstream cosmology explains it as a coincidence and puts it down to a precarious balance between expansion and galaxy formation on the one hand and rate of ionization on the other. For lower redshifts, expansion and galaxy formation have the effect of reducing the density of H1 clouds but the density of quasars also reduces, producing a reduction in the local background UV which reduces the rate at which the clouds disappear by ionization under the set column density. And if the quasars are not at their redshift distances it would change the redshift dependence of the results. But the fact alone of the quasars not being at their redshift distances would significantly change our understanding of modern cosmology. 

\section{Conclusion}

\begin{table*}[t]
\small
\begin{center}
\caption{\label{tab:table1}Straw poll on how the evidence stacks up.$^{i}$ }
\end{center}
\begin{tabular*}{\textwidth}{@{\hspace{\tabcolsep}
\extracolsep{\fill}}lcccc}
\hline\hline
&Evidence 										&Pro   			&Con 				&Comments \\
\hline
&Hubble law 									& $\times$ 	& 					&Derived from general relativity \\
&SNe Ia distance modulus 			& $\times$ 	& 					&Some selection bias, or, intergalactic dust \\
&Dark energy 									&  					& $\times$	&Required from SNe Ia but physically unknown \\
&Dark matter 									&  					& $\times$	&Required from SNe Ia but physically unknown \\
&SN 1997ff time dilation 			& $\times$ 	& 					&Evidence against no time dilation \\
&SN metallicity vs. redshift							&  					& $\times$		&Contrary to expectation \\
&Quasar proper motion					&  					& $\times$		&If verified, very bad for  $\Lambda$CDM model \\
&Quasar metallicity vs. redshift				&  					& $\times$		&Contrary to expectation \\
&Quasar luminosity variations	&						& $\times$ 		&Explained with luminosity evolution\\
&GRB luminosity variations		& $\times$	& $\times$		&Explained with luminosity evolution\\
&Angular size vs. redshift		&						& $\times$		&Explained with ad hoc size evolution\\
&Surface brightness vs. redshift&					& $\times$		&Explained with ad hoc size evolution\\
&Galaxy size vs. redshift		 	&$\times$		& $\times$		&Unexplained by merger theory\\
&Galaxy clusters at high redshift & & $\times$ &Unexplained `too big, too early'\\
&Existence of CMBR	 					& $\times$ 	&							&Predicted in 1948 but first observed in 1941\\
&CMBR shadowing by clusters		&						& $\times$	 	&Results from SZE otherwise unexplained\\
&CMBR temperature vs. redshift& $\times$	& $\times$		&Inconsistent results within the same cloud\\
&Quasars and Ly$\alpha$ absorbers& $\times$	&							&Doubts from MgII  absorbers toward GRBs	 		\\
&GRBs and MgII absorbers		 	&						& $\times$		&`Astonishing' nearly all aligned\\
&Gunn-Peterson trough	 				& $\times$	&							&Doubts on redshift of ``era of reionization''\\
&H1 cloud spacing vs. redshift& $\times$	& $\times$		&Evidence for both expansion and static\\
\hline
\end{tabular*}
\vskip 2mm
$^{i}$This is not intended to be definitive as the relative weights for evidences are not assigned.

\end{table*}

The best evidence in support of an expanding cosmos is the type Ia supernova observations. However, to choose the candidate supernovae, the standard concordance model is used. And yet those same observations can be made to fit a static universe without the time dilation factor necessary to the big bang universe. In this case the main line of evidence in support of the big bang is the $(1+z)$ time dilation factor but if that is due to a selection effect then there is no definitive evidence for an expansion as required.
  
And why do quasars, supposedly the most distance sources in the Universe, not show any evidence of the required cosmological time dilation? The Universe could simply be static--that would neatly solve the problem. Or the quasars may not be so distant--not at their redshift distances. But to save the standard model, one must assume that there has been a conspiracy of competing effects, including an accumulation of black hole mass at the core of these quasars, over cosmic time, that exactly cancels any observable time dilation. 

The Hubble diagram fits a static universe with a simple Euclidean non-expanding space just as well as it does the standard concordance big bang model. In the former case no dark matter, no dark energy, no inflation--all unknown in the lab--are needed. The former extrapolates the simple Hubble law to all redshifts. And it should be realized that there have been suggested many alternatives \citep{Marmet} for the mechanism behind the observed redshifts that don't require cosmological expansion however very little research has been expended on such.

Nevertheless a mechanism for cosmic redshifts (the Hubble law) has been neatly derived from Einstein's general theory, which has been successfully tested in the solar system and with pulsar binary pairs. The latter test the theory in different domains to that of cosmological redshifts, yet adds support that the same theory would apply elsewhere. 

Looking at the angular sizes of galaxies as a function of redshift the static universe model provides a better fit than the standard model and with the least number of assumptions. However, by suitably choosing, ad hoc, evolution in size of galaxies as a function of redshift (by orders of magnitude more than any observation) the standard model can be saved. There is some recent evidence on the growth of individual high redshift galaxies with stars that supersized themselves by gobbling up surrounding fuel, creating stars up to 100 times the mass of our sun. Other than this, the size evolution of galaxies in the standard model by mergers is a difficult research problem.

Taking together all the evidences presented here (see Table I), in my opinion, it is impossible to conclude either way whether the Universe is expanding or static. The evidence is equivocal; open to more than one interpretation.   It would seem that cosmology is far from a precision science, and there is still a lot more work that needs to be done to resolve the apparently contradictory evidence.


\begin{thebibliography}{999}

\bibitem[\protect\citeauthoryear{Aguirre \& Zoltan}{2008}]{Aguirre}  Aguirre, A., \& Zoltan, A., (2000) Astrophys. J., 532, 28.
\bibitem[\protect\citeauthoryear{Aprile et al.}{2010}]{Aprile2010} Aprile E., et al. (XENON100 Collaboration) (2010)  Phys. Rev. Lett., 105, 131302.
\bibitem[\protect\citeauthoryear{Aprile et al.}{2011}]{Aprile2011} Aprile E., et al. (XENON100 Collaboration) (2011) \\  arXiv/1104.2549.
\bibitem[\protect\citeauthoryear{Andrews}{1999}]{Andrews}  Andrews, T.B., (1999) ASP Conf. Series, Vol. 193, A.J. Bunker, and W.J.M. van Breugel (Eds.), Astron. Soc. of Pacific, S. Francisco, 407.
\bibitem[\protect\citeauthoryear{Arp}{1987}]{ArpQRC} Arp, H., Quasars, Redshifts and Controversies, Interstellar Media, Cambridge University Press, Berkeley, CA, 1987.
\bibitem[\protect\citeauthoryear{Arp}{1998}]{ArpSR} Arp, H., Seeing Red, Redshifts, Cosmology and Academic Science, Apeiron, Montreal, 1998.
\bibitem[\protect\citeauthoryear{Arp \& Russell}{2001}]{Arp2001} Arp, H., \& Russell, D. (2001) Astrophys. J., 549, 802.
\bibitem[\protect\citeauthoryear{Arp \& Fulton}{2008}] {ArpFulton} Arp, H., \& Fulton, C., (2008) arXiv/0802.1587
\bibitem[\protect\citeauthoryear{Ashmore}{2009}]{Ashmore2009}  Ashmore, L., (2009) ASP Conf. Series, Vol. 413, F. Potter (Ed.),  Astron. Soc. of Pacific, S. Francisco, 3.
\bibitem[\protect\citeauthoryear{Becker et al.}{2001}]{Becker}  Becker, R. H., Fan, X., and White, R. L. (2001) Astron. J., 122, 2850.
\bibitem[\protect\citeauthoryear{Bell}{2002}]{Bell2002} Bell, M.B. (2002) Astrophys. J., 566, 705; Bell, M.B. (2002)  Astrophys. J., 567, 801.
\bibitem[\protect\citeauthoryear{Bennett et al.}{2003}]{Bennett2003}  Bennett, C. L., Halpern, M., Hinshaw, G. et al. (2003) Astrophys. J. Suppl., 148, 1. 
\bibitem[\protect\citeauthoryear{Bielby \& Shanks}{2007}]{Bielby}  Bielby, R.M. \& Shanks, T. (2007) M.N.R.A.S., 382, 1196. 
\bibitem[\protect\citeauthoryear{Bloom et al.}{2003}]{Bloom2003} Bloom, J.S., Frail, D.A., and Kulkarni, S.R. (2003) Astrophys. J., 594, 674.
\bibitem[\protect\citeauthoryear{Borgonovo}{2004}]{Borgonovo2004} Borgonovo, L. (2004) Astron. \& Astrophys., 418, 487.
\bibitem[\protect\citeauthoryear{Burgay et al.}{2003}]{Burgay2003} Burgay, M., et al. (2003) Nature, 426, 531.
\bibitem[\protect\citeauthoryear{Carrera \& Giulini}{2006}]{Carrera2006} Carrera, M. \& Giulini, D. (2010) Rev. Mod. Phys., 82, 169.
\bibitem[\protect\citeauthoryear{Chang}{2001}]{Chang2001} Chang, H-Y. (2001) Astrophys. J.,  557, L85.
\bibitem[\protect\citeauthoryear{Chang et al.}{2002}]{Chang2002} Chang, H-Y., Yoon, S-J., and Choi, C-S.  (2002) Astron. \& Astrophys., 383, L1.
\bibitem[\protect\citeauthoryear{Cooperstock et al.}{1998}] {Cooperstock} Cooperstock, F.I., Faraoni, V.,  and Vollick, D.N. (1998) Astron. J., 503,  61.
\bibitem[\protect\citeauthoryear{Crawford}{2009}]{Crawford2009}  Crawford, D.F., (2009) arXiv/0901.4169
\bibitem[\protect\citeauthoryear{Crawford}{2011}]{Crawford2011}  Crawford, D.F., (2011) J. of Cosmology, in press, \\ \url{journalofcosmology.com/crawford1.pdf}
\bibitem[\protect\citeauthoryear{Das Gupta}{1988}]{DasGupta}  Das Gupta, P., Narlikar, J.V.,  and Burbidge, G.R. (1988) Astron. J., 95, 5.
\bibitem[\protect\citeauthoryear{Davis et al.}{2003}]{Davis2003} Davis, T.M., Lineweaver, C.H., and Webb, J.K. (2003) Am. J. Phys. 71, 358.
\bibitem[\protect\citeauthoryear{Fan et al.}{2001}]{Fan} Fan, X., Narayanan, V.K., Lupton, R.H., et al. (2001) Astron. J.,  122, 2833.
\bibitem[\protect\citeauthoryear{Foley et al.}{2005}]{Foley}  Foley, R.J., Filippenko, A.V.,  Leonard, D.C.,  Riess, A.G.,   Nugent, P., and Perlmutter, S. (2005) Astrophys. J., 626, L11. 
\bibitem[\protect\citeauthoryear{Galianni et al.}{2005}]{Galianni} Galianni, P., Burbidge, E.M., Arp, H., Junkkarinen, V., Burbidge, G. and Zibetti, S. (2005) Astrophys. J.,  620, 88.
\bibitem[\protect\citeauthoryear{Gallagher et al.}{2008}]{Gallagher2008} Gallagher, J.S.,  Garnavich, P.M., Caldwell, N, Kirshner, R.P., Jha, S.W., Li, W., Ganeshalingam, M. and Filippenko, A.V. (2008) Astrophys. J., 685, 752.
\bibitem[\protect\citeauthoryear{Gamow}{1948}]{Gamow} Gamow, G. (1948) Phys. Rev., 74, 505.
\bibitem[\protect\citeauthoryear{Ge et al.}{1997}]{Ge}  Ge, J., Bechtold, J., and Black, J. H. (1997) Astrophys. J., 474, 67.
\bibitem[\protect\citeauthoryear{Greggio}{2010}]{Greggio2010} Greggio, L. (2010) M.N.R.A.S., 406, 22.
\bibitem[\protect\citeauthoryear{Goldhaber at al.}{2001}]{Goldhaber}  Goldhaber, G., Groom, D.E., Kim, A., et al. (2001) Astrophys. J., 558, 359.
\bibitem[\protect\citeauthoryear{Goobar et al.}{2002}]{Goobar}  Goobar, A., Bergstr\"om, L., and M\"ortsell, E. (2002)  Astron. \& Astrophys., 384, 1.
\bibitem[\protect\citeauthoryear{Hawkins}{2001}]{Hawkins2001}  Hawkins, M.R.S. (2001)  Astrophys. J., 553, L97. 
\bibitem[\protect\citeauthoryear{Hawkins}{2007}]{Hawkins2007}  Hawkins, M.R.S. (2007)  Astron. \& Astrophys., 462, 581.
\bibitem[\protect\citeauthoryear{Hawkins}{2010}]{Hawkins2010}  Hawkins, M.R.S. (2010)  M.N.R.A.S., 405, 1940.
\bibitem[\protect\citeauthoryear{Howell et al.}{2009}]{Howell2009} Howell, D.A., Sullivan, M., Brown, E.F., et al. (2009) Astrophys. J., 691, 661.
\bibitem[\protect\citeauthoryear{Hoyle}{1959}]{Hoyle}  Hoyle, F. (1959) Paris Symposium on Radio Astronomy (IAU Symp. 9, URSI Symp. 1), R.N. Bracewell (Ed.), Stanford University Press, Stanford (CA), 529.
\bibitem[\protect\citeauthoryear{Hoyle et al.}{2011}]{Hoyle2011} Hoyle, B., Jimenez, R., and Verde, L. (2011) Phys. Rev. D., 83, 103502. 
\bibitem[\protect\citeauthoryear{Hubble}{1929}]{Hubble1929} Hubble, E. (1929) Proc. of the National Academy of Sciences, 15, 168. 
\bibitem[\protect\citeauthoryear{Hubble \& Tolman}{1935}]{Hubble1935}  Hubble, E.P., \& Tolman, R.C. (1935) Astrophys. J., 82, 302.
\bibitem[\protect\citeauthoryear{Hubble}{1936}]{Hubble1936} Hubble, E.P. (1936) Astrophys. J., 84, 517.
\bibitem[\protect\citeauthoryear{Hubble}{1947}]{Hubble1947} Hubble, E.P. (1947) Publ. Astron. Soc. Pac., 59, 153.
\bibitem[\protect\citeauthoryear{Hu et al.}{2002}]{Hu}  Hu, E.M., Cowie, L.L., McMahon, R.G., Capak, P., Iwamuro, F., Kneib, J.-P., Maihara, T., and Motohara, K. (2002) Astrophys. J., 568, L75.
\bibitem[\protect\citeauthoryear{Hulse \& Taylor}{1975}]{Hulse1975}  Hulse, R.A. \& Taylor, J.H.  (1975) Astrophys. J., 201, L55.
\bibitem[\protect\citeauthoryear{Ivanov}{2001}]{Ivanov2001}  Ivanov, M.A., (2001) Gen. Rel. and Grav., 33, 479; ERRATUM, (2003) 35, 939.
\bibitem[\protect\citeauthoryear{Ivanov}{2010}]{Ivanov2010}  Ivanov, M.A., No-time-dilation corrected Supernovae Ia and GRBs data and low-energy quantum gravity, Contribution to the VI Int. Workshop on the Dark side of the Universe (DSU2010), Guanajuato U., Leon, Mexico, 1-6 June, 2010. \url{ivanovma.narod.ru/no-time-dilation10.html}
\bibitem[\protect\citeauthoryear{Kapahi}{1987}]{Kapahi}  Kapahi, V.K. (1987) Observational Cosmology (IAU Symp. 124), A. Hewitt, G. Burbidge, and L. Z. Fang (Eds.), Reidel, Dordrecht, 251.
\bibitem[\protect\citeauthoryear{Kirkman et al.}{2007}]{Kirkman} Kirkman, D., Tytler, D., Lubin, D., Charlton, J. (2007) M.N.R.A.S., 376, 1227.
\bibitem[\protect\citeauthoryear{Kormendy et al.}{2010}]{Kormendy} Kormendy, J., Drory, N., Bender, R., and Cornell, M.E. (2010) Astrophys. J., 723, 54.
\bibitem[\protect\citeauthoryear{Kowalski et al.}{2008}]{Uniondata}  Kowalski, M., Rubin., D., Aldering, G., et al. (2008) Astrophys. J., 686, 749.
\bibitem[\protect\citeauthoryear{Lamagna et al.}{2007}]{Lamagna} Lamagna, L., Battistelli, E.S., De Gregori, S., De Petris, M., Luzzi, G., Savini, G. (2007) New Astronomy Reviews, 51, 381.
\bibitem[\protect\citeauthoryear{Lanzetta et al.}{1999}] {Lanzetta}  Lanzetta, K.M., Chen, H.-W., Pascarelle, S., Yahata, N., and Yahil, A. (1999) ASP Conf. Series, Vol. 193, A. J. Bunker, and W.J.M. van Breugel (Eds.), Astron. Soc. of Pacific, S. Francisco, 544.
\bibitem[\protect\citeauthoryear{La Violette}{1986}] {LaViolette1986} La Violette, P.A. (1986) Astrophys. J., 301, 544.
\bibitem[\protect\citeauthoryear{Lerner}{2006}]{Lerner2006}  Lerner, E.J. (2006) AIP Conf. Proc., Vol. 822, E.J. Lerner and J.B. Almeida (Eds.), AIP, 60.
\bibitem[\protect\citeauthoryear{Lerner}{2009}]{Lerner2009}  Lerner,  E.J. (2009) ASP Conf. Series,  Vol. 413, F. Potter (Ed.), Astron. Soc. of Pacific, S. Francisco, 12. 
\bibitem[\protect\citeauthoryear{Lieu et al.}{2006}]{Lieu2006}  Lieu, R., Mittaz, J.P.D., and Zhang, S-N. (2006) Astrophys. J., 648, 176.
\bibitem[\protect\citeauthoryear{Lieu}{2007}]{Lieu2007} Lieu, R. (2007) arXiv/0705.2462
\bibitem[\protect\citeauthoryear{L\'opez-Corredoira}{2003}]{Lopez-Corredoira2003}  L\'opez-Corredoira, M. (2003) Observational Cosmology: caveats and open questions in the standard model, in Book Recent. Res. Devel. Astronomy \& Astrophys., Vol. 1, 561.
\bibitem[\protect\citeauthoryear{L\'opez-Corredoira \& Guti\'errez}{2007}]{Lopez-Corredoira2007}  L\'opez-Corredoira, M. and Guti\'errez, C.M. (2007) Astron. \& Astrophys., 461, 59.
\bibitem[\protect\citeauthoryear{L\'opez-Corredoira}{2010}]{Lopez-Corredoira2010}  L\'opez-Corredoira, M. (2010) Int. J. of Mod. Phys. D, 19, 245. 
\bibitem[\protect\citeauthoryear{Lyne et al.}{2004}]{Lyne2004} Lyne, A.G., et al., (2004) Science, 303, 1153.
\bibitem[\protect\citeauthoryear{McKellar}{1941}]{McKellar}  McKellar (1941) A., Pub. of the Dominion Astrophysical Obs. 7(15), 251. 
\bibitem[\protect\citeauthoryear{MacMillan}{2005}]{MacMillan} MacMillan, D.S. (2005) ASP Conf. Series, Vol. 340, J. Romney and M. Reid (Eds), Astron. Soc. of Pacific, S. Francisco, 477.
\bibitem[\protect\citeauthoryear{Marmet}{2011}]{Marmet}  Marmet, L. (2011) \url{www.marmet.org/cosmology/redshift/mechanisms.pdf}
\bibitem[\protect\citeauthoryear{Miralda-Escud\'e et al.}{2000}]{Miralda-Escude}  Miralda-Escud\'e, J., Haehnelt, M., and Rees, M. J. (2000) Astrophys. J., 530, 1.
\bibitem[\protect\citeauthoryear{Molaro et al.}{2002}]{Molaro}  Molaro, P, Levshakov, S.A., Dessauges-Zavadsky, M., and D'Odorico, S. (2002) Astron. \& Astrophys., 381, L64.
\bibitem[\protect\citeauthoryear{Norris}{2002}]{Norris2002} Norris, J.P. (2002) Astrophys. J., 579, 386.
\bibitem[\protect\citeauthoryear{Peacock}{1999}]{Peacock1999} Peacock, J.A. (1999) Cosmological Physics, (Cambridge: Cambridge University Press), pp. 87–89.
\bibitem[\protect\citeauthoryear{Perlmutter et al.}{1999}]{Perlmutter}  Perlmutter, S., et al. (1999)  Astrophys. J., 517, 565.
\bibitem[\protect\citeauthoryear{Piran}{2004}]{Piran2004} Piran, T. (2004) Rev. Mod. Phys., 76, 1143.
\bibitem[\protect\citeauthoryear{Prochter et al.}{2006}] {Prochter}  Prochter, G.E., et al. (2006) Astrophys. J., 648, L93. 
\bibitem[\protect\citeauthoryear{Rauch}{1998}]{Rauch}  Rauch, M. (1998) Annu. Rev. Astron. Astrophys., 36, 267
\bibitem[\protect\citeauthoryear{Reynaud \& Jaekel}{2008}]{Reynaud2008} Reynaud, S. \& Jaekel, M-T. Notes of a lecture given during the International School of Physics Enrico Fermi on Atom Optics and Space Physics (Varenna, July 2007), arXiv/0801.3407.
\bibitem[\protect\citeauthoryear{Riess et al.}{1997}]{Riess1997}  Riess, A.G., Filippenko, A.V., Leonard, D.C., et al. (1997) Astron. J., 114, 722.
\bibitem[\protect\citeauthoryear{Riess et al.}{1998}]{Riess1998}  Riess, A.G., Filippenko, A.V., Challis, P., et al. (1998) Astron. J., 116, 1009.
\bibitem[\protect\citeauthoryear{Rowan-Robinson}{2002}]{Rowan-Robinson}  Rowan-Robinson, M. (2002)  M.N.R.A.S., 332, 352.
\bibitem[\protect\citeauthoryear{Scarpa et al.}{2007}]{Scarpa}  Scarpa, R., Falomo, R., and Lerner, E.J. (2007)  Astrophys. J., 668, 74.
\bibitem[\protect\citeauthoryear{Schaefer}{2007}]{Schaefer2007} Schaefer, B.E. (2007) Astrophys. J., 660, 16.
\bibitem[\protect\citeauthoryear{Schilling}{2006}]{Schilling}  Schilling, G. (2006) Science, 313, 749.
\bibitem[\protect\citeauthoryear{Shim et al.}{2011}]{Shim} Shim, H., Chary, R-R., Dickinson, M., Lin, L., Spinrad, H., Stern, D., Yan, C-H. (2011)  Astrophys. J., in press, arXiv/1103.4124
\bibitem[\protect\citeauthoryear{Shen \& Song}{2003}]{Shen2003} Shen, R-F. \& Song, L-M. (2003) Publ. Astron. Soc. of Japan, 55, 345.
\bibitem[\protect\citeauthoryear{Simon et al.}{2007}] {Simon} Simon, L.E., Hamann, F.W., Pettini M. (2007) Rev. Mex. Astron. A. (Serie de Conferencias),  29, 177.
\bibitem[\protect\citeauthoryear{Songaila et al.}{1994}]{Songaila}  Songaila, A., Cowie, L.L., Vogt, S., et al. (1994) Nature, 371, 43.
\bibitem[\protect\citeauthoryear{Srianand et al.}{2000}]{Srianand} Srianand, R., Petitjean, P., and Ledoux, C (2000) Nature, 408, 931.  
\bibitem[\protect\citeauthoryear{Steinhardt \& Caldwell}{1998}]{Steinhardt}  Steinhardt, P.J. \& Caldwell, R.R. (1998) ASP Conf. Ser., Vol. 151, Astron. Soc. of Pacific, S. Francisco, 13.
\bibitem[\protect\citeauthoryear{Stott et al.}{2011}]{Stott} Stott, J.P., Collins, C.A., Burke, C., Hamilton-Morris, V., and Smith, G.P. (2011) M.N.R.A.S., 414, 445.
\bibitem[\protect\citeauthoryear{Talbot \& Varshni}{1999}] {Talbot} Talbot, J.\& Varshni, Y.P., Proper Motion of the quasar Ton 202, American Astronomical Society, 194th AAS Meeting, \#73.16; (1999) Bull. Am. Astron. Soc., 31, 952.
\bibitem[\protect\citeauthoryear{Tejos et al.}{2009}]{Tejos}  Tejos, N., et al.  (2009)  Astrophys. J., 706, 1309. 
\bibitem[\protect\citeauthoryear{Trujillo et al.}{2006}]{Trujillo}  Trujillo, I., F\"orster Schreiber, N.M., Rudnick, G., et al. (2006) Astrophys. J., 650, 18.  
\bibitem[\protect\citeauthoryear{Turner}{1999}]{Turner}  Turner, M.S. (1999) ASP Conf. Ser., Vol. 165, Astron. Soc. of Pacific, S. Francisco, 431.
\bibitem[\protect\citeauthoryear{Varshni}{1982}] {Varshni}  Varshni, Y.P. (1982) Speculations in Science and Technology, 5, 521.
\bibitem[\protect\citeauthoryear{Wang \& Han}{2010}]{Wang2010} Wang, B. \& Han, Z. (2010) Astron. \& Astrophys., 515, A88
\bibitem[\protect\citeauthoryear{Zwicky}{1929}]{Zwicky} Zwicky, F. (1929) Proc. Natl Acad. Sci. USA, 15, 773. 


\end{thebibliography}
\end{document}